\RequirePackage{ifpdf}
\ifpdf 
\documentclass[pdftex]{sigma}
\else
\documentclass{sigma}
\fi

\begin{document}

\renewcommand{\PaperNumber}{014}

\FirstPageHeading

\ShortArticleName{On Chaotic Dynamics in Rational Polygonal Billiards}

\ArticleName{On Chaotic Dynamics in Rational Polygonal Billiards}

\Author{Valery B. KOKSHENEV} 
\AuthorNameForHeading{V.B. Kokshenev}

\Address{Departamento de Fisica, Universidade Federal de Minas
Gerais, Instituto de Ciencias Exatas, Caixa Postal 702, CEP 30123-970, Belo
Horizonte, MG, Brazil}

\Email{\href{mailto:valery@fisica.ufmg.br}{valery@fisica.ufmg.br}} 

\ArticleDates{Received June 23, 2005, in final form October 29, 2005;
Published online November 13, 2005}

\Abstract{We discuss the interplay between the piece-line regular and
vertex-angle singular boundary effects, related to integrability and chaotic
features in rational polygonal billiards. The approach to controversial
issue of regular and irregular motion in polygons is taken within the
alternative deterministic and stochastic frameworks. The analysis is
developed in terms of the billiard-wall collision distribution and the
particle survival probability, simulated in closed and weakly open polygons,
respectively. In the multi-vertex polygons, the late-time wall-collision
events result in the circular-like regular periodic trajectories (sliding
orbits), which, in the open billiard case are likely transformed into the
surviving collective excitations (vortices). Having no topological analogy
with the regular orbits in the geometrically corresponding circular
billiard, sliding orbits and vortices are well distinguished in the weakly
open polygons via the universal and non-universal relaxation dynamics.}


\Classification{37D45; 37D50; 51E12; 60C05; 60J60}

\section{Introduction}

Classical polygonal billiards, formed by the piecewise-linear billiard
boundary with the vertex angles that are rational multiplies of $\pi $, are 
\emph{rational polygons}, known as \emph{non-chaotic }systems (see \cite{G96} for review). 
They are therefore quite distinct from the \emph{Sinai billiard}~\cite{Sin79} 
(SB) and the \emph{Bunimovich billiard}~\cite{Bun79} (BB), in
which classical \emph{chaotic} motion regimes are due to, respectively,
dispersive effects caused by the boundary, formed by the disk and the
square, and the interplay between boundary segments, formed by the
semi-circles and the square. Moreover, rational polygons of $m$ equal sides
and $m$ equal vertices (hereafter, $m$-gons~\cite{G96}) exhibit chaotic-like
changes in the associated quantum-level spectra \cite{CC89}, but the
fluctuations found \cite{SS95} are very close to those known in the
universal Gaussian statistics. These controversial evidences for the chaotic
and regular behavior of polygons are a challenge to dynamic theory of
billiards. In the present paper we develop a physical insight into the
problem, based on analysis of the simulation data on the billiard collision
statistics. Within a more general context, the delicate problem of the
interplay between regular and irregular segments, which constitute the total
billiard boundary, is ultimately related to apparent controversy between the
causality and randomness. Our approach to the problem is developed through
the alternative deterministic and stochastic frameworks, which elucidate the
duality of chaotic and non-chaotic features in the intrinsic dynamics of the
rational polygons.

The singular effects, provoked by the vertex-angle splitting of orbits, is
one of the major problem in dynamics of non-dispersed billiards \cite{G96}.
Being countable, they cannot be described within the standard Markov
partition scheme. Similarly to the recent study on the arc-touching effects
in dispersing SB \cite{tobe}, we give analysis of vertex-splitting effects
on the basis of the genera\-lized diffusion equation. The issue of our
investigation is the late-time dynamics memory effects induced by sides and
vertices in both closed and the open rational polygons.

\section{Collision statistics}

\subsection{Collision distribution function}

In billiards formed by classical particles of unit mass, moving with unit
velocities, the billiard \emph{collision distribution function\ }$D(n,t)$ is
determined as the probability for a given particle to undergo $n$ random
collisions with the boundary in a time interval $t$ (for the rigorous
definition of $D(n,t)$ and for the way of its numerical observation, see 
\cite{GG94}). With accounting for the property of ergodicity, this provides
the \emph{mean collision number } 
\begin{gather}
n_{\rm c}(t)\equiv\langle n\rangle_{\rm c}=\int_{0}^{\infty}nD(n,t)dn=\int n(\boldsymbol{x},t)\,
d\mu(\boldsymbol{x})=\frac{t}{\tau_{\rm c}}, \qquad t\gg\tau_{\rm c}.   \label{n1}
\end{gather}
Here $n(\boldsymbol{x},t)$ is the number of billiard collisions during time $t$,
for a particle of the position set $\boldsymbol{x}=(x,y,\theta)$, which includes
its location $(x,y)$ and the \emph{velocity launching angle} $\theta=[0,2\pi]
$ counted of the $x$-axis of the billiard table in the space $\Omega$. The
continues-time evolution of the particle orbits preserves the Liouville
measure \cite{Che97} 
\begin{gather}
d\mu(\boldsymbol{x})=\frac{1}{2\pi A}dxdy\,d\theta.   \label{Lm}
\end{gather}
and establishes the \emph{mean collision time} \cite{GG94,Che97,Gar97}
\begin{gather}
\tau_{\rm c}=\frac{\pi A}{P},   \label{tau-c}
\end{gather}
controlled by the accessible area $A$ and the perimeter $P$ of a billiard
table. Such a description of the continuos dynamic system can be derived
from the late time behavior of the associated collision subsystem \cite{GG94,Che97,VK01}.

The case of $m$-gon is specified by circumscribing of the rational polygon
below a circle of radius $r$. For a given $m$, the billiard area 
$A_{m}=(mr^{2}/2)\sin(2\pi/m)$ and the perimeter $P_{m}=2mr\sin(\pi/m)$
provide the mean collision time 
\begin{gather}
\tau_{\rm cm}=\frac{\pi r}{2}\cos\left(\frac{\pi}{m}\right).   \label{tau-cm}
\end{gather}
In the `infinite-vertex' $m$-gon limit, one naturally arrives at the 
\emph{circle-billiard} (CB) of radius $r$, with the mean collision time 
\begin{gather}
\tau_{{\rm c}\infty}=\tau_{\rm c}^{\rm (CB)}=\frac{\pi r}{2}.   \label{tau-inf}
\end{gather}
A general question arises how geometrical changes of the billiard boundary
are followed by corresponding classical dynamics. This question is related
to the problem of the existence of the correspondence principle in polygonal
billiards~\cite{Man00}. In view of equation (\ref{tau-inf}), it sounds as
whether the $\infty$-gon is a dynamic analog of the integrable CB.

\subsection{Dynamic diffusion regimes}

\begin{figure}[t]
\begin{minipage}[t]{7.5cm}
\centerline{\includegraphics[width=7.5cm]{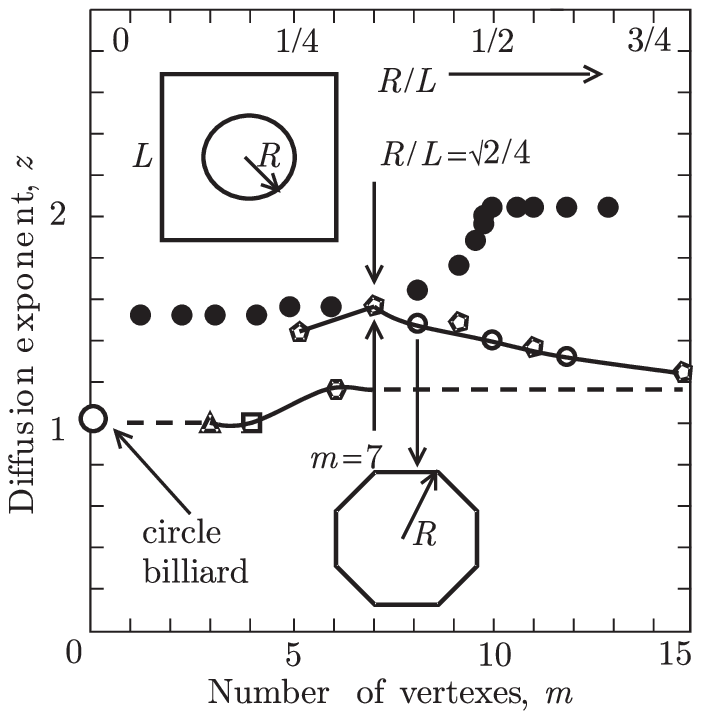}}
\vspace{-3mm}
\caption{Observation of the diffusion dynamics in the Sinai (above) and
polygonal (below) billiards. \emph{Points:} closed circles --- numerical data
for the SBs of side $L=1$ for distinct dispersing disks of radii~$R$. The
open triangle, square, pentagon, \textit{etc.} are the data for the regular
polygons with $m=3,4,5$, \textit{etc.} equal sides. The lines are a guide to
the eye. The experimental error does not exceed the size of symbols.}
\end{minipage}
\hfill 
\begin{minipage}[t]{7.5cm}
\centerline{\includegraphics[width=7.5cm]{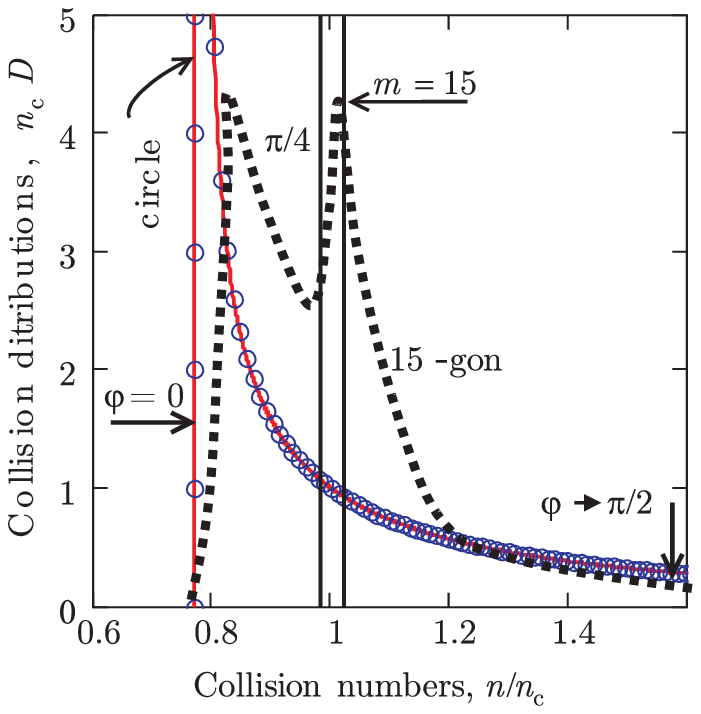}}
\vspace{-3mm}
\caption{Observation of the collision distribution function. The \emph{solid
lines} are theoretical predictions for the CB (\ref{D-CB}) and the 
$15$-gonal billiard, estimated in the deterministic-motion approximation. The
open circles and the point-dotted line schematically represent numerical
simulation of the late-time distributions. For the original data see Fig.~1
in~\cite{RC02}.}
\end{minipage}

\end{figure}

We have communicated \cite{RC02} that the chaotic disk-dispersing SBs and
non-chaotic vertex-splitting polygonal billiards display a similar behavior
attributed to the enhanced diffusion dynamics. In polygons, this late-time
dynamics is associated with the singular arc-splitting effects, produced by
singularities of orbits provoked by the vertices~\cite{G96}. The 
\emph{stochastic approach} proposed for analysis of these effects is based on
exploration of the generalized diffusion equation, which manifests the
dynamic relaxation of non-Markovian time evolution~\cite{MK00}. This
fundamental model-independent equation, which generalizes the Brownian
diffusion, is given for the varian\-ce of particle displacements 
$\Delta^{2}r(t)$ realized during time $t$. In application to $m$-gons, this
equation was introduced through the variance of the number of random
collisions~\cite{RC02}, namely
\begin{gather}
\Delta ^{2}n_{\rm cm}(t)\equiv \langle (n-n_{\rm cm})^{2}\rangle_{\rm c}\backsim \left( \frac{t}{\tau
_{\rm cm}}\right) ^{2/z_{m}}, \qquad t\gg \tau _{\rm c},  \label{delta-n-m}
\end{gather}
represented with the help of equations (\ref{n1}) and (\ref{tau-cm}). Here $%
z_{m}$ stands for the boundary-dependent \emph{diffusion exponent} (which
for the Brownian diffusion is equal to 2). In the case of chaotic SB, of
side $L$ and of disk of radius~$R$, the random walk treatment of the
problem provided the equation~\cite{tobe} 
\begin{gather}
\Delta ^{2}n_{\rm c}^{\rm (SB)}(t)\equiv \langle (n-n_{\rm c}^{\rm (SB)})^{2}\rangle_{\rm c}\backsim \left[ 
\frac{t}{\tau _{\rm c}(R)}\right] ^{2/z(R)}, \qquad t\gg \tau _{\rm c}(R),
\label{delta-n-SB}
\end{gather}
corresponding to equation (\ref{delta-n-m}), taken with the mean collision
time $\tau _{\rm c}(R)$, defined in equation~(\ref{tau-c}), and the diffusion
exponent $z(R)$. Equations (\ref{delta-n-m}) and (\ref{delta-n-SB}) form a
basis for the observation of distinct stationary regimes, driven by the
billiard geometry. The uniform initial conditions were simulated by the
standard random number 
generator RAN1\footnote{For details of the numerical experiments, see~\cite{VK01}.\label{fot1}} and the results
of the analysis of temporal behavior of the directly observed variances for
the billiard-wall number collisions are presented in Fig.~1.

In SB, the universal ($R$-independent) late-time diffusive regime is due to
the stabilization of particle-displacement statistics with $z^{(\exp
)}(R)=1.50\pm0.05$. This was rationalized~\cite{tobe} through the universal
behavior of the particle propagator function~\cite{MK00} and the L\'{e}vy
flights, which occur between the two distant scatterers, belonging to
infinite Bleher's corridors of the dynamically equivalent Lorentz Gas (LG) 
\cite{Ble92}. This almost-free motion along the principal diagonal and
non-diagonal corridors, induced by arc-touching dispersive events,
terminates at $R\geq L\sqrt{2}/4$, when the diagonal corridor becomes closed.

In $m$-gons, the non-universal dynamic regimes are well distinguished of
those, observed in the chaotic SB and the non-chaotic CB, having the unique 
\emph{ballistic regime} defined by $z^{\rm (SB)}(0)=z^{\rm (CB)}=1$. In the
unilateral-triangle and square polygonal billiards, the vertex-splitting
effects are fairly weak, because the regular orbits are due to the particles
mainly move along the infinite trajectories, well known in the case 
$m=3$~\cite{Vee89}. As seen in Fig.~1, these particles expose their ballistic
dynamics via $z_{3}^{(\exp)}\thickapprox z_{4}^{(\exp)}\thickapprox 1$,
characteristic of the integrable~CB. In contrast, the enhanced diffusion in 
\emph{pentagon} and \emph{heptagon}, developed by the vertex splitting of
regular orbits, achieves its maximal level, indicated by the highest
exponents $z_{5}^{(\exp)}\thickapprox z_{7}^{(\exp)}=3/2$. In what follows,
the observed distinct late-time dynamics is examined through the qualitative
description of long-living orbits.

\subsection{Orbit topology}

It is disturbing that the question of whether every polygon has a periodic
orbit remains open~\cite{G96}. Meanwhile, any orbit, which is perpendicular
to a side returns to this side parallel to itself and therefore it is
periodic \cite{Rui91,Trou04}. The existence of the \emph{perpendicular
periodic trajectories} in rational polygons was proved independently in 
\cite{Bosh92} and \cite{GSV92}. The stability of the corresponding periodic
orbits in arbitrary polygon was also shown \cite{Trou04}.

In the integrable CB and in `almost integrable' $m$-gons, limited by, say, $%
m=3,4$ and $6$, the launching angle $\theta $ (\ref{Lm}) is almost good
integral of motion for the particles randomly injected on the billiard$^{\ref{fot1}}$. 
Consequently, the orbit-set classification of the late-time living orbits
in such non-chaotic billiards can be introduced through the \emph{collision
angle} $\varphi =[0,\pi /2]$, estimated from the normal to the boundary wall 
\cite{VK01,KN00}. This motion integral provides the basis for our 
\emph{deterministic approach} to billiard dynamics proposed in~\cite{RC02} and
then developed for the square billiard~\cite{tobe} and polygons~\cite{KV03}.

In CB, all regular orbits are ranged between the \emph{diameter-beating} set
and the `\emph{whispering-gallery}' sets established, respectively, by the
collision angles $\varphi=0$ and $\varphi\thickapprox\pi/2$ (shown in
Fig.~1). In triangle billiards, the existence of the `\emph{perpendicular
orbit}' with $\varphi=0$ was suggested in \cite{Rui91}, and then proved for
any $m$-gon~\cite{Bosh92,GSV92}. In the case of square billiard, these
orbits correspond to the well-known `\emph{bouncing-ball}' orbits 
\cite{VK01,BB90,Pik92}, given by $\varphi\thickapprox0$- angle set and the 
$\varphi\thickapprox\pi/2$-set, moving in the horizontal and vertical
directions, associated with the horizontal and vertical free-motion
corridors. In the even-$m$-gons, the marginal vertex-to-vertex particle
trajectories~\cite{G96} formally correspond to the stable `diameter-beating'
orbits in CB. In contrast, the stable \emph{sliding} \emph{orbits}, observed
for $m>4$, are treated as an analog of the whispering-gallery orbits in CB
responsible for the late-time superdiffusive dynamics, revealed in Fig.~1 in
the chaotic-like pentagon ($z_{5}^{(\exp)}=1.42$) and in the ordered-like 
\emph{hexagon} ($z_{6}^{(\exp)}=1.18$). The chaotic effects are maximal in 
\emph{heptagon} ($z_{7}^{(\exp)}=1.48$), but with further increase the
number of vertices, when $m\geq8$, they attenuate gradually. The analysis
shown in Fig. 1 suggests that in the `infinite-vertex' limit, the
superdiffusive motion is not transformed into the ballistic regime, i.e., 
$z_{\infty}>1$. Although the geometrical correspondence exists, as it is
exposed in equation (\ref{tau-inf}), the sliding orbits are topologically
distinct from the whispering-gallery orbits in the CB, and therefore the
late-time dynamic correspondence between the $\infty$-gon and the CB does
not exist. This statement is additionally tested in the next section.

\subsection{Observation of orbits}

In CB of radius $r$, the collision distribution function%
\begin{gather}
D^{\rm (CB)}(n,t)=\frac{\pi^{2}n_{\rm c}^{3}}{16n^{4}}\left( 1-\frac{\pi^{2}n_{\rm c}^{2}
}{16n^{2}}\right) ^{-1/2}, \qquad n_{\rm c}^{\rm (CB)}=\frac{2t}{\pi r} 
\label{D-CB}
\end{gather}
defined in equation (\ref{n1}), is found through the Liouville measure.
Similarly, if one ignores the vertex-splitting (irregular) effects a given $m$-gon, 
the particle-collision function, corresponding to the regular
orbits, can be obtained through the model (equivalent-side) approximation 
\cite{KV03}, namely 
\begin{gather}
D_{m}^{\rm (reg)}(n,t) =\frac{\sin\varphi_{\rm cm}}{n_{\rm cm}\varphi_{\rm cm}^{2}}
\left[ 1-\left( \frac{n}{n_{\rm cm}}\frac{\sin\varphi_{\rm cm}}{\varphi_{\rm cm}}\right)
^{2}\right] ^{-1/2}, \qquad n_{\rm cm}=\frac{t}{\tau_{\rm cm}},  \nonumber \\
\mbox{for}\quad \varphi_{\rm cm}\cot\varphi_{\rm cm}  \leq
n/n_{\rm cm}\leq\varphi_{\rm cm}/\sin\varphi_{\rm cm}, \quad \mbox{otherwise} \quad D_{m}^{\rm (reg)}=0,
\label{Dm}
\end{gather}
where $\varphi_{\rm cm}=\pi/2m$, for the \emph{odd} $m$-gon, and $\varphi
_{\rm cm}=\pi/m$ , for the \emph{even} $m$\emph{-gon}. A comparative analysis
between the distributions observed in $15$-gon and CB is displayed
schematically in Fig.~2.

In Fig.~2, the diameter-beating orbit ($\varphi=0$) and the
whispering-gallery orbit ($\varphi\lessapprox\pi/2$) are well observable in
CB. In the case of $15$-gon, the vertex-singular effects, manifested by
deviations from the regular-orbit behavior, seem tend to be adapted through
the bouncing-ball-like ($\varphi\approx0$) and the \emph{sliding orbits} 
($\varphi\approx\pi/2$). We see that both kinds of orbits with $%
\varphi\approx\pi/2$ (shown by open circles and solid squares) are well
distinguished, because, the sliding orbits are topologically distinct from
their prototype in CB. This finding is supported by the distinct diffusion
exponents, $z_{15}^{(\exp)}=1.2$ and $z^{\rm (CB)}=1$, analyzed in Fig.~1.

In a given $m$-gon, the characteristic collision time 
\begin{gather}
t_{\rm cm}^{\rm (reg)}(\varphi)=\tau_{{\rm c}\infty}
\frac{\sin(\varphi_{m})\cos(\pi /m)}{\varphi_{m}\cos(\varphi-\psi_{m})},   \label{tc-m}
\end{gather}
was also found \cite{RC02,KV03}, in the regular-orbit approximation. 
Here~$\psi_{m}=$ $\varphi_{\rm cm}$ and $\psi_{m}=0$, for, respectively, the 
\emph{even} $m$\emph{-gon} and \emph{odd }$m$\emph{-gon} cases, discussed in
equation (\ref{Dm}). One can see that $t_{\rm cm}(\varphi)\gg\tau_{\rm cm}$ for
the set $\varphi\thickapprox\pi/2$. The longest-living regular (sliding)
orbits are therefore expected to be observed in the weakly open $m$-gons.

\section{Surviving dynamics}

In the case of weakly open $m$-gons, one deals with $N(0)=10^{6}$ uniformly
distributed point particles, which are allowed to escape through a small
opening in the billiard boundary$^{\ref{fot1}}$. A~temporal behavior of 
$N(t) $ non-escaped orbits (particles) is commonly scaled by the \emph{mean
escape time} \cite{GG94,VK01,BB90,Alt96} 
\begin{gather}
\tau_{\rm e}=\tau_{\rm c}\frac{P}{\Delta} \quad \mbox{with}\quad \Delta\ll P, 
\label{tau-e}
\end{gather}
where the mean collision time $\tau_{\rm c}$ is given in equation (\ref{tau-c}).
The late-time asymptote of the billiard \emph{survival probability}
\begin{gather}
S(t)=\frac{N(t)}{N(0)}\propto\left( \frac{\tau_{\rm e}}{t}\right) ^{\delta }, 
\quad \mbox{for}\quad t\gg\tau_{e},   \label{S-def}
\end{gather}
describing the algebraic-behavior through the \emph{decay dynamic exponent} 
$\delta$. In \emph{non-chaotic} square billiard, the algebraic decay with the
exponents $\delta\lessapprox1$ was first reported in~\cite{BB90}. A careful
study of decay dynamics in the \emph{integrable} CB and in the `almost
integrable' $4$-gon established \cite{VK01} two distinct channels of slow
relaxation. The first channel is due to the regular-orbit motion with the
decay exponent $\delta=1$, provided by the whispering-gallery orbits, and
the \emph{secondary channel} originates from the bouncing-ball orbits,
exposed \cite{VK01} by $\delta<1$.

In \emph{chaotic} closed and weakly open classical systems (including
Hamiltonian systems), exemplified by BB \cite{Pik92,Alt96,Bun85},
infinite-horizon SB \cite{KN00,FS95}, and by corresponding low-density 
LG~\cite{Ble92,FM84}, the overall algebraic decay was found numerically through
the geometry-dependent exponents $\delta \geq 1$. The channel of 
algebraic- type relaxation with $\delta =1$, common for both chaotic 
\cite{KN00,Alt96,FS95} and non-chaotic~\cite{VK01,BB90} billiards, seems to be
generic for all incompletely hyperbolic systems with smooth convex
boundaries. Its independence with respect to the billiard spatial dimension 
\cite{FS95}, its insensitivity to the details of boundary shape \cite{VK01},
including the location of a small opening \cite{Alt96}, and to the initial
conditions \cite{VK01} suggests that the late-time $\alpha $-relaxation in
classical systems is the \emph{universal primary relaxation}. Unlike the
primary relaxation, the \emph{secondary relaxation }in chaotic billiards
with $\delta >1$, is shown to be rather sensitive to the billiard geometry 
\cite{VK01}, as well as to the billiard table dimension $d$ \cite{FS95}, and
to the initial conditions \cite{VK01}. Example is the dynamic-exponent
constraint $1<\beta \leq d$ estimated \cite{FS95} and observed in both
chaotic BB \cite{Alt96} and SB \cite{FS95,ACG96}. As a result, the
dynamic-regime classification can be proposed for the open classical
billiards in terms of the decay dynamic exponent (\ref{S-def}), namely 
\begin{gather}
\mbox{primary channel:} \quad \delta =\alpha =1, \quad \mbox{universal regime;}
\nonumber \\
\mbox{secondary channel:}\quad  \delta =\beta \left\{ 
\begin{array}{ll}
<1, & \mbox{non-chaotic regime}, \\ 
>1, & \mbox{chaotic regime}.
\end{array}\right.   \label{class}
\end{gather}

\begin{figure}[t]
\begin{minipage}[t]{7.5cm}
\centerline{\includegraphics[width=7.5cm]{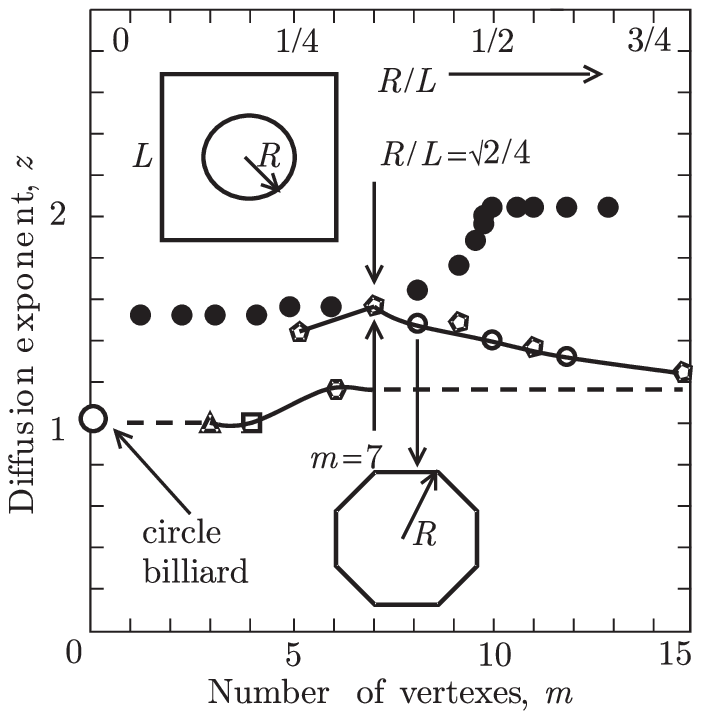}}
\end{minipage}
\hfill
\raisebox{1mm}[0pt][0pt]{\parbox[b]{7.5cm}{\caption{Analysis of the temporal evolution of the survived orbits in
rational polygons with small opening width $\Delta =0.05R$. The observation
times are reduced through the escape charac\-te\-ristic time $\tau _{\rm e}=300$,
chosen common for all cases. \emph{Points}: numerical data for the simulated 
$m$-gons (squares) and the corresponding CB (circles).} }}

\vspace{-2mm}
\end{figure}

In Fig.~3 we study the decay dynamics in $m$-gons with $m\geq 8$ through the
number of survived particles (\ref{S-def}) $N_{m}(t)=S_{m}(t)N_{m}(0)$. For
the particular case of small opening $\Delta =0.05R$, the universal
relaxation (\ref{class}) remains stable until $m=64$, but when $m\geq
m_{\alpha }^{(\exp )}=128$ the primary-relaxation observation window becomes
closed. The sliding regular-orbit dynamics likely transforms into a 
\emph{singular-orbit} chaotic-like motion, when intensively affected by vertices.
This is indicated by the dynamic decay exponent $\beta _{m}^{(\exp )}=1.2$
and equation~(\ref{class}). Also, the experimental observations was elaborated
for the large opening $\Delta =0.20R$, when $\beta _{m}^{(\exp )}=1.1$ and $%
m_{\alpha }^{(\exp )}=32$ were derived in this case (see Fig.~3 in \cite{KV03}). 
We examine that the upper limit for the $\alpha $%
-channel-observation window shows its sensitivity to the opening width.
\smallskip This suggests that the primary relaxation dominates in $m$-gons
with $3\leq m<m_{\alpha }^{(\exp )}$, where the regular-orbit decay motion
is established by the universal decay exponent $\delta _{m}=\alpha =1$. The
latter is due to the `\emph{perpendicular}' and sliding orbits, which
exhibit, respectively, the non-chaotic and chaotic behavior, which are not
distinguished within this channel. In order to prove that the observed
chaotic-like regime in the open system is really caused by the long-living
sliding orbits, we deduce the observation conditions for such a decay
dynamics.

Similarly to the boundary case $R=L\sqrt{2}/4$, corresponding to the
universal-to-non-universal crossover in relaxation dynamics of the chaotic
SB (shown in Fig.~1), the $\alpha$-to-$\beta$ crossover occurs in $m$-gons
with $m=$ $m_{\alpha}$ and is associated with the transformation of the
particle motion, induced by regular segments of boundary, into that provoked
by its singular parts. Just as the periodic motion in the crystalline
lattice is observed through the collective excitations (phonons), so the
exploration by particles of translational periodicity of the billiard table
with large $m$ can provide the \emph{collective-particle motion} of
characteristic time $t_{\rm cm}^{\rm (coll)}(\varphi)=\tau_{{\rm c}\infty}\cos^{-1}(\varphi)$, 
permitted for any regular orbit, preserving the
collision angle $\varphi$. This qualitative estimate, following from
equation~(\ref{tc-m}) taken at $m\gg1$, indicates that the \emph{idealized}
sliding orbits, observed now at $\varphi\thickapprox\pi/2$ and $m\gg 1$,
leave the piecewise-linear part of the billiard boundary. The real
sliding-orbit \emph{excitations} of finite \emph{relaxation times} 
$\tau_{\rm cm}^{\rm (slide)}$, are thought as marginal regular orbits
established by the maximal collision angles $\varphi _{m}^{\rm (slide)}=\varphi_{m}$ 
(\ref{tc-m}) with $\tau_{\rm cm}^{\rm (slide)}=t_{\rm cm}^{\rm (coll)}(\varphi_{m})$. 
In contrast to the ideal sliding orbits, we also
determine the \emph{ideal vortex orbits} as sliding along the
piecewise-linear part of the boundary without reflection. Such irregular
`orbits' are formally distinguished by the collision angles $\varphi _{m}^{\rm (vort)}
=\Phi_{m}/2$, where $\Phi_{m}$ are rational vertex angles. In
view of the $m$-fold rotational symmetry of the billiard table, the
existence of the real singular-like \emph{vortex excitations} (or vortices)
is expected from the preservation of the angular momentum for a certain set
of vertex-angle correlated orbits. We therefore assume that the real
long-living sliding (regular) orbits are precursors of the vortex-like
(singular) orbits, attributed to the relaxation times $\tau_{\rm cm}^{\rm (vort)}
=t_{\rm cm}^{\rm (coll)}(\Phi_{m}/2)$. In this way, the $\alpha$-to-$\beta$%
-relaxation crossover is attributed to the observed regular-to-singular
sliding-orbit transformation, that occurs starting from the large numbers 
$m_{\alpha}^{(\exp)}$. The observation condition for the crossover between
the two distinct real dynamic regimes can be approximated by the condition
of \ disappearance of the ideal sliding orbits above $m_{\alpha}$ and that
of the ideal vortex orbits below $m_{\alpha}$.

A small opening of width $\Delta$ ($\ll P_{m}$ ) can be arranged at any
point of boundary \cite{Alt96}. The favorable survival situation in the $%
\alpha $-relaxation regime, induced by the regular part of the boun\-dary of
length $L_{m}=P_{m}/m$, must exclude the vertex-angle effects under the
constraint $m<m_{\alpha}$. The geometrical condition, at which the ideal
vortices effectively escape from the billiard table, corresponds to the
opening of width $\Delta\ll L_{m}$, located at one of the angle vertices.
Conversely, in the late-time $\beta$-relaxation regime, the regular orbits
do not survive at all, if any side is absorbed by the opening, i.e.\
when $\Delta\gg L_{m}$, with $m>m_{\alpha}$. Hence, the $\alpha$-to-$\beta$
crossover in relaxation is qualitatively ensured by the condition $%
\Delta=L_{m}$ at $m=m_{\alpha}$. Taking into consideration that in polygons
with large number of sides $P_{m}\approx P_{\infty}=2\pi r$, one obtains
the desirable criterion of the $\alpha$-to-$\beta$ crossover (\ref{class}),
namely 
\begin{gather}
m_{\alpha}=\frac{2\pi r}{\Delta}.   \label{m-alfa}
\end{gather}
The experimental data $m_{\alpha}^{(\exp)}=128$ and $32$, discussed above,  
are in agreement with the estimates $m_{\alpha}=126$ and $31$,
following from equation (\ref{m-alfa}). This finding justifies our surmise
that the observed long-living vortices in $\beta$-relaxation are due to the
sliding orbits modified by intensive vertex splitting. Finally, as follows
from equation (\ref{m-alfa}) in the limit $\Delta\rightarrow0$ and as
experimentally supported in Fig.~1, the observation window for vortices
disappears in closed polygons.

\section{Conclusion}

We have rationalized the late-time relaxation dynamics, observed in rational
polygons, in terms of the so-called \emph{sliding orbits}, associated with
the periodic trajectories. The dynamic properties of the corresponding
orbits, determined by the collision-angle set close to $\pi/2$, are shown to
be due to the interplay between the piece-line (regular) part of the
billiard boundary and its vertex-angle (singular) part. The sliding orbits
are beyond the simplified polygonal-orbit classification by Gutkin~\cite{G96}, 
where the regular orbits are presented by the
`infinite-past-to-infinite-future' trajectories, and the `mild
discontinuity', caused by vertex-splitting effects, is due to the
`infinite-past-to-vertex', `vertex-to-infinite-future', and
`vertex-to-vertex' marginal trajectories. Also, they are in a certain way opposed to the 
 \emph{perpendicular orbits}, rigorously studied in 
\cite{Trou04,Bosh92,GSV92}. The experimental prove for the existence of the
long-living periodic sliding trajectories challenges to further mathematical
research in polygons.

The sliding orbits are well pronounced in simulated dynamics in both closed
and open rational polygons. When the number of vertices is small ($m<5$),
these orbits are ineffective in the dynamics driving by the long-living
bouncing-ball-like orbits. With increasing of the number of sides (and
vertices), the sliding orbits become stable and therefore regular, but
simultaneously expose chaotic-like features, in a way similar to that in the
chaotic SB. These features have been revealed here through the diffusion
exponent and the collision distribution function. The chaotic-like sliding
orbits, nevertheless, play the role of regular whispering-gallery
orbits, known in the integrable CB.

In the open $m$-gons, the chaotic properties of the long-living sliding
orbits remain masked and hidden, when observed via the primary relaxation
within the domain $8\leq m<m_{\alpha}(\Delta)$. This situation changes above
the $\alpha $-to-$\beta$ crossover, where new collective excitations,
denominated by \emph{vortices}, become to be pronounced in the secondary
billiard relaxation. These chaotic excitations are associated with the
collective vertex-splitting effects, established in multi-vertex $m$-gons,
which number of vertices exceeds $m_{\alpha}=2\pi r/\Delta$.

Regardless the proposed rationalization of the regular-like sliding orbits
and the irregular-like vortices, we have experimentally revealed that the
boundary-collision events, observed in the collective dynamics in rational
polygons are dual of regular and singular of isolated particle collisions
with the corresponding parts of the billiard boundary. Being related to the
zero-measure singularities in phase space, these effects violate the
integrability of polygons~\cite{G96}, as well as the classical-to-quantum
correspondence principle~\cite{RC02}. The latter finding is due to the
memory of vertex-splitting effects, which do not disappear in the billiard
dynamics when $m\rightarrow\infty$, and thereby forbid the interchange of
temporal ($t\rightarrow\infty$) and the spatial ($m\rightarrow\infty$)
limits.

\subsection*{Acknowledgments}

The financial support of the Brazilian agency CNPq is acknowledged.

\LastPageEnding

\end{document}